\documentclass[aps,prl,showpacs,twocolumn,groupedaddress]{revtex4}
\usepackage{graphicx}

\begin{document}

\title{Freak waves in the linear regime: A microwave study}

\author{R. H\"{o}hmann$^1$}
\author{U. Kuhl$^1$}
\author{H.-J. St\"{o}ckmann$^1$}
\author{L. Kaplan$^2$}
\author{E. J. Heller$^3$}
\affiliation{$^1$Fachbereich Physik der Philipps-Universit\"at Marburg, D-35032 Marburg, Germany\\
$^2$Department of Physics, Tulane University, New Orleans, Louisiana 70118, USA\\
$^3$Department of Physics and Department of Chemistry and Chemical Biology, Harvard University, Cambridge, Massachusetts 02138, USA}

%PACS: 42.25.Dd Wave propagation in random media
%PACS: 05.45.Mt Quantum chaos; semiclassical methods
%PACS: 42.25.Bs Wave propagation, transmission and absorption

\date{\today}
\begin{abstract}
Microwave transport experiments have been performed in a quasi-two-dimensional resonator with randomly distributed conical scatterers. At high frequencies, the flow shows branching structures similar to those observed in stationary imaging of electron flow. Semiclassical simulations confirm that caustics in the ray dynamics are responsible for these structures. At lower frequencies, large deviations from Rayleigh's law for the wave height distribution are observed, which can only partially be described by existing multiple-scattering theories. In particular there are ``hot spots'' with intensities far beyond those expected in a random wave field. The results are analogous to flow patterns observed in the ocean in the presence of spatially varying currents or depth variations in the sea floor, where branches and hot spots lead to an enhanced frequency of freak or rogue wave formation.
\end{abstract}

\maketitle

In this Letter, we present a microwave transport study through
a scattering system composed of randomly placed metallic cones, each mimicking an $r^{-2}$
potential on the scale of its radius. For wavelengths smaller than or comparable to the scatterer size, we find branching
structures that are reminiscent of electron current
distributions seen in two-dimensional electron systems~\cite{top01}.
At wavelengths larger than the cone size,
the bulk of the intensity distribution approaches a multiple-scattering correction to Rayleigh statistics, as expected in
multiple-scattering media~\cite{nie95}.
However, the probability of finding very
high intensities is still greatly enhanced. Even
larger fluctuations are observed after Fourier transforming the frequency-domain measurements
into the time-domain; the extreme time-domain events may be compared to freak wave
events in the ocean~\cite{hel08}. The enhanced probability
of large intensities at longer wavelength is a residual of
presumably stronger enhancements at shorter wavelength.

Scanning tunneling microscopy studies of electron flow injected through point contacts in a high-mobility two-dimensional electron gas (2DEG) by Topinka et\,al.~\cite{top01} exhibited intricate branching patterns of fractal appearance. This behavior was in contrast to the simple random wave prediction for the probability distribution of wave intensities, $p_{\rm Rayleigh}(I)=e^{-I}$, where $I\sim|\psi|^2$ is normalized to one, and $|\psi|$ is the wave height. Topinka et\,al. showed that the evolution of caustics (singularities of the ray density) in a random potential (with a given strength and correlation length) is responsible for the branching patterns.
Kaplan later showed that the probability distribution of the branches may be computed analytically~\cite{kap02}.
The phenomenology holds equally well for the evolution of wave patterns in the sea. In shallow water, wave focusing may be caused by depth variations in the sea floor, and may lead to amplification of tsunami waves~\cite{dob06, ber07}. In deep water, the effect of eddy currents has been studied by Heller et\,al.~\cite{hel08}, who showed that even after accounting for dispersion in wavelength and direction, random currents greatly increase the likelihood of large amplitude events; they argued that such events might act as a trigger for nonlinear instability effects in freak wave formation. Thus, without diminishing the importance of nonlinear processes, an understanding of the linear regime is essential for the proper understanding of freak wave physics.

\begin{figure}[ht]
  \includegraphics[width=.7\columnwidth]{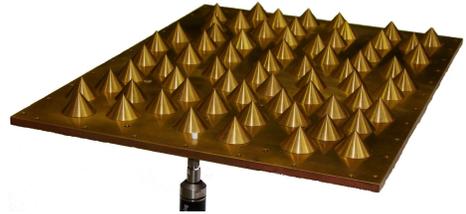}
  \caption{\label{fig:setup}
  \noindent (Color online) Photograph of one of the two scattering arrangements used.
  The platform has width 260\,mm and length 360\,mm. Each cone has diameter 25 \,mm and height 15\,mm. The probe antenna is fixed in a horizontally movable top plate located 20\,mm above the bottom (not shown).
  }
\end{figure}

Motivated primarily by the 2DEG experiments~\cite{top01}, we undertook a microwave experiment to study the transport of waves through an arrangement of randomly distributed scatterers. Figure~\ref{fig:setup} shows a photograph of the experimental setup. The metallic bottom plate supports the scattering arrangements made up of about 55 to 60 brass cones. The source antenna is mounted close to one of the short sides, and varying its position enables the incoming waves to arrive from different directions. The drain antenna is mounted in the top plate (not shown), and acts as a weak probe. The top plate can be moved in both horizontal directions, allowing for a spatial mapping of the wave fields within the scattering arrangement.

For quasi-two-dimensional systems with parallel top and bottom plates separated by a vertical distance $d$ (without scatterers), the electromagnetic wave equations reduce to a single, scalar equation for the
perpendicular component $E_\perp(x,y,z)$ of the electric field. Furthermore, we can write $E_\perp(x,y,z)=E(x,y)\cos(n\pi z/d)$, where $n$ is the transverse quantum number. This results in a two-dimensional wave equation for $E(x,y)$,
\begin{equation}\label{eq:helmholtz}
  \left[-\frac{\partial^2}{\partial x^2}-\frac{\partial^2}{\partial
y^2}+\left(\frac{n\pi}{d}\right)^2\right]E(x,y)=k^2E(x,y)\,,
\end{equation}
which for $n=0$ is equivalent to the free stationary Schr\"odinger equation in the plane~\cite{stoe99}.
The number of active modes $n$ depends on frequency. Eq.~(\ref{eq:helmholtz}) remains approximately true when $d$ varies slowly with position (on the scale of the wavelength). For $n>0$, the additional term then mimics a potential, $V(x,y)=[n\pi/d(x,y)]^2$~\cite{kim05b}. Each cone in the experiment corresponds to a repulsive central potential: $V(r)\propto [C+{\rm min}(r,r_0)]^{-2}$, where $r_0=12.5$\,mm is the cone radius and $C=16.7$\,mm is a constant determined by the cone geometry. At higher orders, however, the height variation leads to a mixing among the modes.

\begin{figure}
  \parbox{0.48\columnwidth}{\includegraphics[width=0.48\columnwidth,clip]{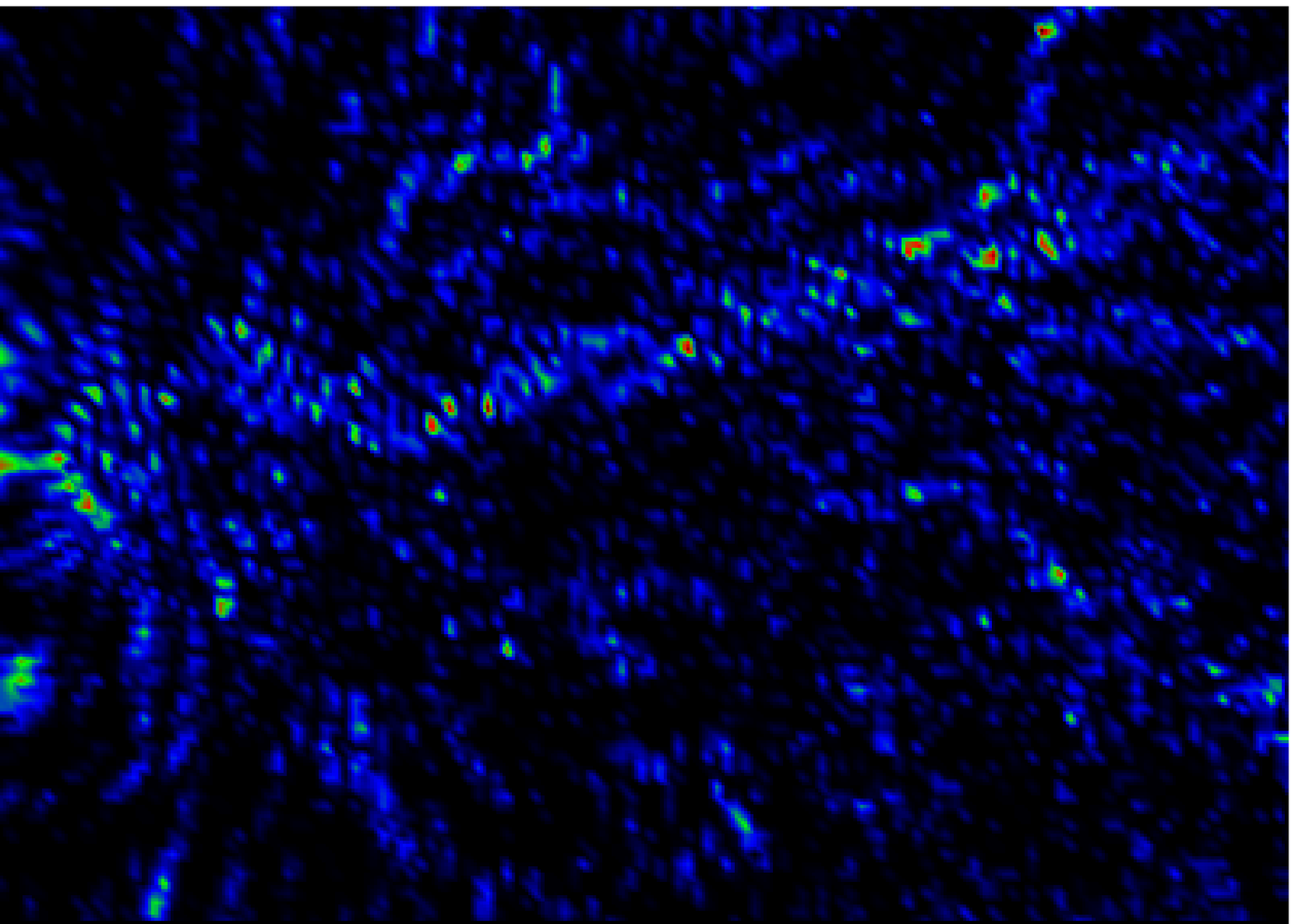}}\hfill
  \parbox{0.48\columnwidth}{\includegraphics[width=0.48\columnwidth,clip]{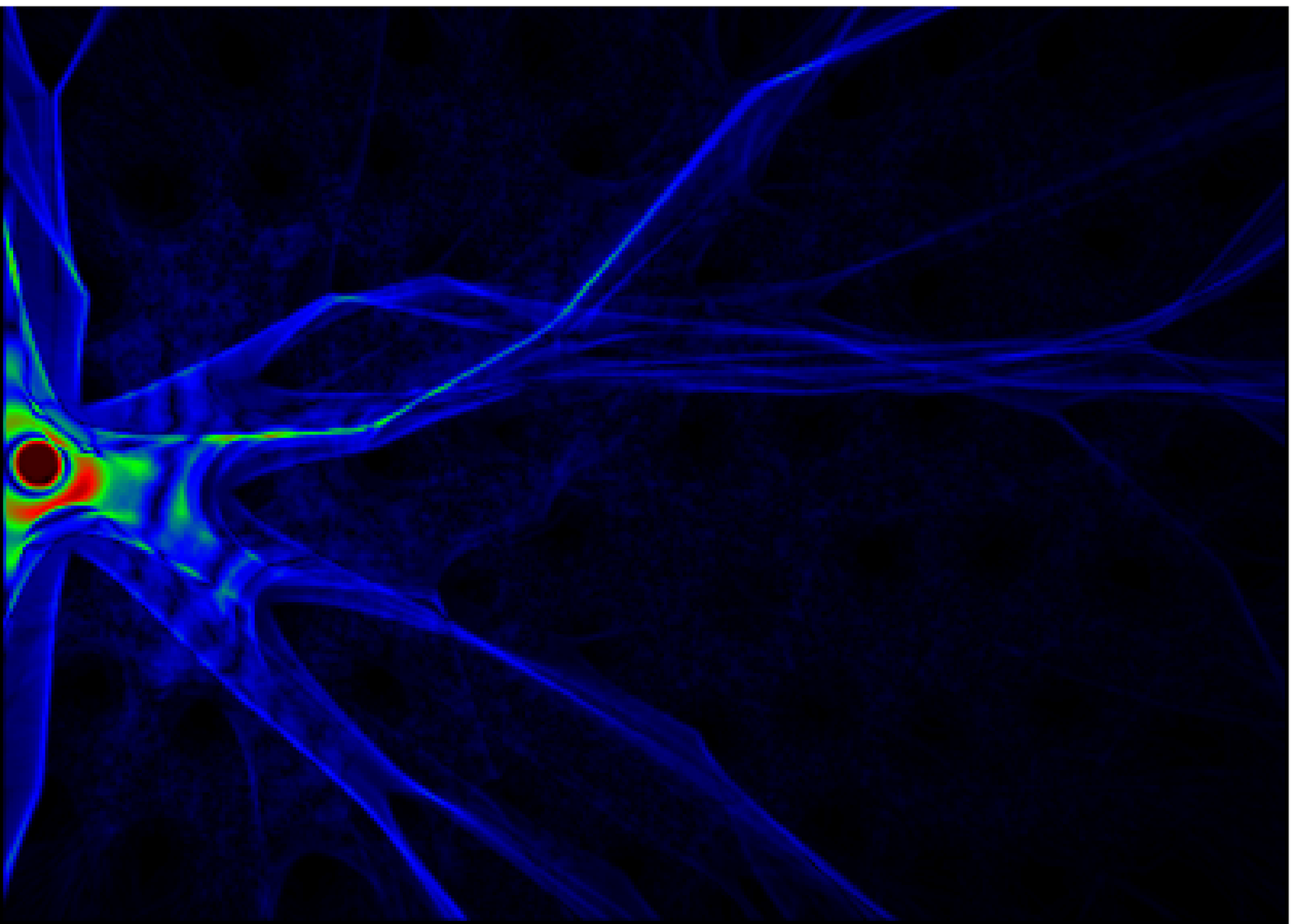}}\\
\caption{ \label{fig:branch}
\noindent (Color online) Comparison of an experimental wave pattern with a classical ray simulation. Left: A wavefunction at frequency $f=30.95$\,GHz. Right: The corresponding semiclassical simulation, with modes 1 through 4 added together.
}
\end{figure}

The left panel of Fig.~\ref{fig:branch} shows a typical wave pattern observed for $f=30$ to $40$ GHz, the upper limit accessible by our equipment. At these frequencies, modes $n=0$ to $4$ are propagating. The wavelengths are between 7.5 and 10\,mm, i.\,e., somewhat smaller than but comparable to the cone diameter. We observe exponential decay of the wave intensity with distance from the source, caused in part by escape of the waves from the scattering setup (since the system is open along the perimeter). This decay is suppressed in all plots. The right panel of Fig.~\ref{fig:branch} shows the results of a ray simulation, obtained by solving the classical equations of motion in the potential generated by the scatterers. The pattern bears a striking similarity with the branchlike structures found by Topinka et\,al.~\cite{top01} in studies of 2DEG electron flow. The present results provide strong evidence in favor of the conjecture~\cite{top01,kap02} that random potentials correlated on the scale of a wavelength are responsible for these features.

To avoid mixing of up to five different modes, most of the experiments have been performed between 7.5 and 15~GHz, where only the first two modes $n=0$ and 1 are propagating. In this regime, however, the wavelengths are large compared to the scatterer size. Hence the separation of the $z$ component in the wave equation is no longer justified, and the interpretation of the cones in terms of a classical particle potential becomes invalid.

In the weakly disordered regime, Nieuwenhuizen and van Rossum~\cite{nie95} calculated the first perturbative correction to Rayleigh's law, yielding $p_{\rm pert}(I) \propto e^{-I}[1+(I^2-4I+2)/3g]$ for the intensity distribution, where $g$ is the dimensionless conductance, exactly what had been found in a microwave transport study by Genack and Garcia~\cite{gen93}. A nonperturbative expression has been calculated by Mirlin et\,al.~\cite{mir98} for the transmission between two antennas embedded in a quasi-one-dimensional disordered sample,
\begin{equation}\label{eq:mirlin}
  p_{\rm dis}(I) \approx \exp\left\{-\frac{\gamma}{2}\left[\ln^2\left(\sqrt{1+\frac{2I}{\gamma}}+\sqrt{\frac{2I}{\gamma}}\right)\right]\right\}\,,
\end{equation}
where $\gamma$ is proportional to the conductance and depends also on the position of the two antennas~\cite{mir98}. For $\gamma \gg I$, we have $p_{\rm dis}(I) \approx e^{-I+2I^2/3\gamma}$, where the $I^2$ term gives the first perturbative correction to Rayleigh's law.
We emphasize that all corrections to Rayleigh contained in such multiple-scattering expressions disappear in the limit of short wavelength (high frequency or large $g$). This is not the case with smooth potentials, which semiclassically deflect the flow; indeed the effect of classical caustics becomes more pronounced at short wavelengths~\cite{kap02}.

\begin{figure}
  \centerline{\includegraphics[width=0.82\columnwidth]{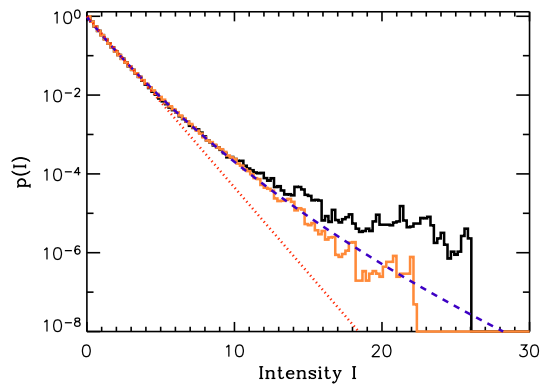}}
  \caption{\label{fig:rayleigh} \noindent (Color online) Probability distribution of intensities. The dark (black) histogram includes all data, while the light (yellow) histogram excludes frequencies associated with the hot spots. The dotted line is the Rayleigh distribution, while the dashed (blue) line is a best fit using the theoretical distribution given by Eq.~(\ref{eq:mirlin})
  ($\gamma \approx 23.5$).
}
\end{figure}

Figure~\ref{fig:rayleigh} shows the intensity distribution found in our experiments, averaged over the complete data set (two scattering arrangements, three source antenna positions for one of
the arrangements, and a frequency range of 7.5 to 11\,GHz). The distribution is well described by Eq.~(\ref{eq:mirlin}) over three orders of magnitude.
Here the situation is comparable to the one found by the Genack group in a number of studies \cite{gen05b}. But for the very high intensities, the probability exceeds
the multiple-scattering theory prediction
by one to two orders of magnitude.

\begin{figure}
  \includegraphics[width=.85\columnwidth]{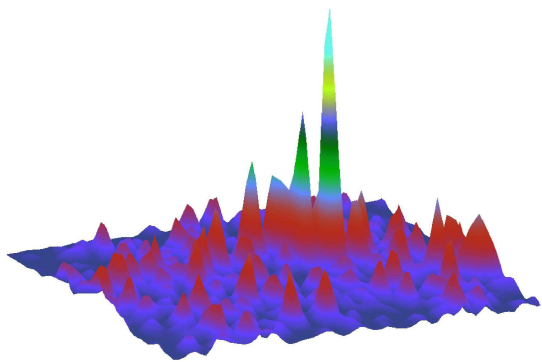}
  \caption{\label{fig:hotspot} \noindent (Color online) A ``hot spot'', observed at a frequency of 8.85\,GHz.
  The experimental probability density for observing such a hot spot is
  one to two orders of magnitude larger than that expected from multiple-scattering theory.
}
\end{figure}

Examining the data, we find that just two regions (and another one at the border of significance) are responsible for these deviations, one of them shown in Fig.~\ref{fig:hotspot}. Each of these ``hot spots'' exists only in a limited frequency window about 500\,MHz in width, and the range of incoming wave directions able to excite each hot spot is only about 20 degrees wide. If the frequency ranges containing these hot spots are omitted from the analysis, the resulting intensity distribution is in full agreement with the multiple-scattering prediction, see Fig.~\ref{fig:rayleigh}.

For the study of time-dependent waves, such as those found in the sea, we must superimpose waves with different frequencies, entering from different directions. To this end we concentrate on one hot spot found at 9.5\,GHz near the center of our scattering arrangement. We vary the antenna position over 80\,mm along the short side of the scattering arrangement (see Fig.~\ref{fig:setup}), and record the resulting field pattern for each antenna position and each frequency. Then time-dependent wave fields are generated by superposition of $N=150$ patterns,
\begin{equation}\label{eq:pulse}
  \psi(\vec{r},t) = \sum_{i=1}^N\psi_{i}(\vec{r})e^{\imath (2 \pi f_i t-\varphi_i)}.
  \label{eq::TransientDef}
\end{equation}
Here $\psi_{i}(\vec{r})$ is a wave pattern at frequency $f_i$, excited with the source antenna at position $x_i$, and $\varphi_i$ is a random phase. The randomly chosen frequencies $f_i$ are normally distributed with the average at $9.5$\,GHz and a standard deviation of $0.764$\,GHz, thus covering approximately the frequency window in which this hot spot is present. Similarly, the antenna positions $x_i$ are taken within the angle of acceptance of the hot spot.

\begin{figure}
  \centerline{
    \includegraphics[width=.88\columnwidth,clip]{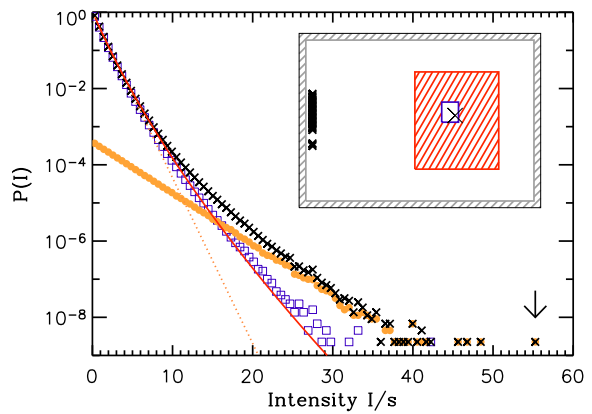}}
  \caption{\label{fg::IntensityDistrTransient} (Color online)
Intensity distribution for the time-dependent wave patterns generated by Eq.~(\ref{eq::TransientDef}) for all measured points [dark (black) crosses], for the hot spot region only [light (orange) solid circles], and for all points excluding the hot spot [light (blue) open squares]. The dotted line is the random wave expectation, while the solid (red) line is given by Eq.~(\ref{eq::kbess}). The arrow indicates the extreme event studied more closely in Fig.~\ref{fig:freak}. The inset shows a sketch of the set up: the dark crosses mark the different exciting antenna positions, the (red) shaded region corresponds to the measured field, and the empty rectangle inside the measured field indicates the hot spot region. The cross inside the hot spot region is the position of the maximal measured intensity.}
\end{figure}

\begin{figure}
  \centerline{\includegraphics[width=.85\columnwidth,clip]{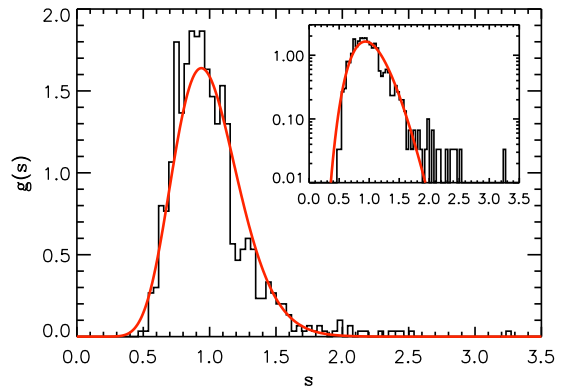}}
  \caption{\label{fg::meanIntensityDistrTransient} (Color online) Distribution of the time-averaged intensities $s$ found for the 780 pixels of our measurement. The inset shows the same data using a semilogarithmic scale. The solid curve is a $\chi^2$ distribution with $\nu=32$ degrees of freedom.}
\end{figure}

Fixing the probe position, we always find a Rayleigh law for the distribution of intensities in a time sequence, $P_{\rm loc}(I)=s^{-1} e^{-I/s}$, but with the time-averaged value $s=\langle I \rangle$ depending on
position. This is nothing but a manifestation of the central limit
theorem. An example taken at the hot spot is given by the light (orange) solid circles in Fig.~\ref{fg::IntensityDistrTransient}. This is the situation a ship experiences at a given
position. If now the ship changes its position, or, alternatively,
if the random currents change, a Rayleigh law
with another $s$ is found. The overall
distribution of time-dependent intensities, collected over position and/or realization of the disorder,
is then given by
\begin{equation}\label{eq:ptot}
  P(I)=\int\limits_0^\infty ds \,s^{-1} g(s)e^{-I/s} \,,
\end{equation}
where $g(s)$ is the probability density to find a local time-averaged intensity $s$.

Normalizing the intensity $s$ to one,
we find that $g(s)$ can be very well described by a chi-square
distribution
\begin{equation}\label{eq::chi2}
  g(s)=\chi_\nu^2(s)=\left(\frac{\nu}{2}\right)^{\frac{\nu}{2}}\frac{1}{\Gamma\left(\frac{\nu}{2}\right)}
  s^{\frac{\nu}{2}-1}\exp\left(-\frac{\nu s}{2}\right) \,,
\end{equation}
where the number of degrees of freedom
$\nu$ increases linearly both with the frequency range used and with the
range of source antenna positions. This is exactly what is expected for a
pulse made up of $\nu/2$ {\it independent} patterns obeying
Rayleigh distributions. A typical example is shown in Fig.~\ref{fg::meanIntensityDistrTransient}. If the input patterns are not Rayleigh distributed, a correspondingly modified $\chi^2$ distribution is obtained. The corrections to expression (\ref{eq::chi2}) become significant in the range of high amplitudes.

With expression (\ref{eq::chi2}) for $g(s)$, the integral (\ref{eq:ptot})
yields the $K$ distribution~\cite{jakemanpusey}
\begin{equation}\label{eq::kbess}
  P(I)=\frac{\nu}{\Gamma\left(\frac{\nu}{2}\right)}\left(\frac{\nu I}{2}\right)^{\frac{\nu}{4}-\frac{1}{2}} K_{\frac{\nu}{2}-1}\left(2\sqrt{\frac{\nu I}{2}}\right)\,,
\end{equation}
where $K_\nu(x)$ is a modified Bessel function. The solid (red) line in
Fig.~\ref{fg::IntensityDistrTransient} is calculated from
Eq.~(\ref{eq::kbess}). It fits nicely with the intensity
distribution found if the hot spot region is excluded (blue squares), but not with
the distribution including the hot spot (black crosses). This is not really a
surprise. Already the inset of Fig.~\ref{fg::meanIntensityDistrTransient} shows that the $\chi^2$
distribution, though generally working well, fails to describe the rare events in the high-intensity tail.

\begin{figure}
  \includegraphics[width=.93\columnwidth]{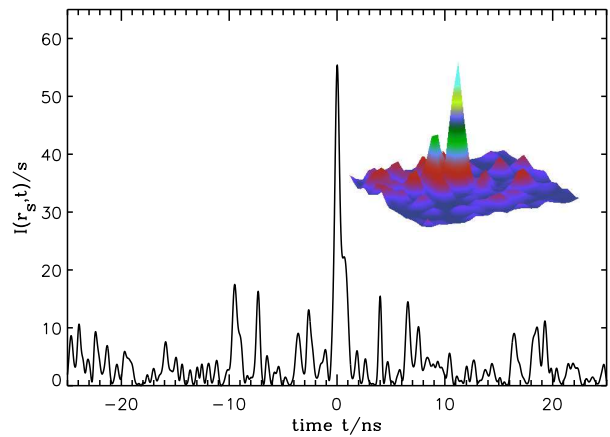}\hfill
  \caption{\label{fig:freak}
  \noindent (Color online) A freak wave event. The time evolution of wave intensity at the center of one of the hot spots is shown for the most extreme event observed. The inset shows the region surrounding the hot spot at the moment of the freak event. A movie of the time evolution in the entire field surrounding the hot spot is available~\cite{movie}.}
\end{figure}

Figure~\ref{fig:freak} shows the most extreme event found in our time series, marked by an arrow in Fig.~\ref{fg::IntensityDistrTransient}, and the inset shows the entire region about the hot spot at the moment of this freak event. In the experiment, we observe events of this magnitude or greater with a probability of $1.3\cdot10^{-9}$. Thus, such events are still quite rare, but the probability is enhanced by 5 orders of magnitude compared with Eq.~(\ref{eq::kbess}), and by 15 orders of magnitude compared to the Rayleigh distribution!

This work has demonstrated the virtues of microwave techniques for obtaining
detailed information on wave transport through disordered surroundings. Simply by varying the frequency, we are able to study both ray dominated branching behavior of flow in a potential landscape, as well as the diffractive multiple-scattering regime. The interpretation of the hot spots in this latter regime has to remain speculative for the moment, in view of the small number of such hot spots showing up in the experiments. However, the narrow angular acceptance of each hot spot, the visually obvious branching behavior at the higher frequencies, and particularly the fact that the observed deviations from Rayleigh statistics get stronger rather than weaker at shorter wavelengths, all support the hypothesis that the hot spots are not resonant wave phenomena but instead are remnants of singularities in the classical dynamics.

\end{document}